\documentclass[runningheads]{svmult}
\usepackage{makeidx}   % allows index generation
\usepackage{graphicx}  % standard LaTeX graphics tool
\usepackage{multicol}  % used for the two-column index
\usepackage{physprbb}  % modified textarea for proceedings,
\makeindex             % used for the subject index
\usepackage{amsmath,amssymb,latexsym}

\begin{document}
\pagestyle{headings}  % switches on printing of running heads
\mainmatter           % start of the contributions
\title*{Jam-avoiding adaptive cruise control (ACC) and its impact on traffic dynamics}
\titlerunning{'Jam-avoiding' adaptive cruise control (ACC)} 
% abbreviated title (for running head)
% also used for the TOC unless \toctitle is used
%
\author{Arne Kesting\inst{1} \and Martin Treiber\inst{1} \and Martin Sch\"onhof\inst{1} \and Florian Kranke\inst{2} \and Dirk Helbing\inst{1,3} }
\authorrunning{A. Kesting, et al.}   % abbreviated author list (for running head)
%
%%%% modified list of authors for the TOC (add the affiliations)
\tocauthor{Arne Kesting, M. Treiber, M. Sch\"onhof, D. Helbing (TU Dresden),
F. Kranke (Volkswagen Wolfsburg)}
\institute{
  Technische Universit\"at Dresden, Institute for Transport \& Economics,\\
  Andreas-Schubert-Strasse 23, D-01062 Dresden, Germany\\
  \and
  Volkswagen AG, Postfach 011/1895, D-38436 Wolfsburg, Germany
  \and
   Collegium Budapest -- Institute for Advanced Study,\\ Szenth\'aroms\'ag u. 2, H-1014 Budapest, Hungary}
\maketitle              % typeset the title of the contribution

%%%%%%%%%%%%%%%%%%%%%%%%%%%%%%%%%%%%%%%%%%%%%%%%%%%%%%%%%%%%%%%%%%%%%%%%%%

\begin{abstract}
Adaptive-Cruise Control (ACC) automatically accelerates or decelerates
a vehicle to maintain a selected time gap, to reach a desired
velocity, or to prevent a rear-end collision. To this end, the ACC
sensors detect and track the vehicle ahead for measuring the actual
distance and speed difference. Together with the own velocity, these
input variables are exactly the same as in car-following models.  The
focus of this contribution is: What will be the impact of a spreading
of ACC systems on the traffic dynamics? Do automated driving
strategies have the potential to improve the capacity and stability of
traffic flow or will they necessarily increase the heterogeneity and
instability? How does the result depend on the ACC equipment level?

We discuss microscopic modeling aspects for human and automated (ACC)
driving.  By means of microscopic traffic simulations, we study how a
variable percentage of ACC-equipped vehicles influences the stability
of traffic flow, the maximum flow under free traffic conditions until
traffic breaks down, and the dynamic capacity of congested
traffic. Furthermore, we compare different percentages of ACC with
respect to travel times in a specific congestion scenario. Remarkably,
we find that already a small amount of ACC equipped cars and, hence, a
marginally increased free and dynamic capacity, leads to a drastic
reduction of traffic congestion.
\end{abstract}

%%%%%%%%%%%%%%%%%%%%%%%%%%%%%%%%%%%%%%%%%%%%%%%%%%%%%%%%%%%%%%%%%%%%%%
%
%%%%%%%%%%%%%%%%%%%%%%%%%%%%%%%%%%%%%%%%%%%%%%%%%%%%%%%%%%%%%%%%%%%%%%
\section{Introduction}
Traffic congestion is a severe problem on European freeways. According
to a study of the European Commission \cite{EU-WhitePaper}, its impact
amounts to 0.5\% of the gross national product and will increase even
up to 1\% in the year 2010. Since building new infrastructure is no
longer an appropriate option in most (Western) countries, there are
many approaches towards a more effective road usage and a more
'intelligent' way of increasing the capacity of the road
network. Examples of advanced traffic control systems are, e.g.,
'intelligent' speed limits, adaptive ramp metering, or dynamic
routing. These examples are based on a centralized traffic management,
which controls the operation and the response to a given traffic
situation. In this contribution, we focus on a local strategy based on
autonomous vehicles, which are equipped with adaptive cruise control
(ACC) systems. The motivation is that a 'jam-avoiding' driving
strategy of these automated vehicles might also help to increase the
road capacity and thus decrease traffic congestion. Moreover, ACC
systems become commercially available to an increasing number of
vehicle types.

An ACC system is able to detect and to track the vehicle ahead,
measuring the actual distance and speed difference. Together with the
own speed, these input data allow the system to calculate the required
acceleration or deceleration to maintain a selected time headway, to
reach a desired velocity, or to prevent a rear-end collision. It
should be emphasized that ACC systems control the longitudinal driving
task. Merging, lane changing or gap-creation for other vehicles still
needs the intervention of the driver. ACC systems promise a gain in
comfort and safety in applicable driving situations, but they are not
yet applied in congested traffic conditions. The next generation of
ACC will successfully extend the application range to all speed ranges
and most traffic situations on freeways including stop-and-go traffic. This leads
to the question: In which way does a growing market penetration of
ACC-equipped vehicles influence the capacity and stability of traffic
flow?  Although there is considerable research on this topic
\cite{Minderhoud}, there is even no clarity up to now about the sign
of the effect. Some investigations predict a positive effect
\cite{Treiber-aut,Davis-ACC}, while others are more pessimistic
\cite{marsden-ACC,kerner-book}.

The contribution is organized as follows: We start with a discussion of
modeling issues concerning the description of human vs. automated
driving and pinpoint the differences between ACC-driven vehicles and
human drivers. In Sec.~\ref{sec:VLA}, we will model three ACC driving
styles, which are explicitly designed to increase the dynamic capacity
and traffic stability by varying the individual driving behavior.
Since the impact on the traffic dynamics could solely be answered by
means of traffic simulations, in Sec.~\ref{sec:results} we perform a
simulation study of mixed freeway traffic with a variable percentage
of ACC vehicles. In Sec.~\ref{sec:diss}, we conclude with a discussion
of our results.

%%%%%%%%%%%%%%%%%%%%%%%%%%%%%%%%%%%%%%%%%%%%%%%%%%%%
%%%%%%%%%%%%%%%%%%%%%%%%%%%%%%%%%%%%%%%%%%%%%%%%%%%%
%
\section{\label{sec:human}Modeling human and automated (ACC) driving behavior}
Most microscopic traffic models describe the acceleration and
deceleration of each individual 'driver-vehicle unit' as a function of
the distance and velocity difference to the vehicle in front and on
the own velocity \cite{Helb-opus,nagel-wagner-vdiff}. Some of these
car-following models have been successful to reproducing the
characteristic features of macroscopic traffic phenomena such as
traffic breakdowns, the scattering in the fundamental diagram, traffic
instabilities, and the propagation of stop-and-go waves or other
patterns of congested traffic. While these collective phenomena can be
described by macroscopic, fluid-dynamic traffic models as well
\cite{GKT}, microscopic models are more appropriate to cope with the
heterogeneity of mixed traffic, e.g., by representing individual
driver-vehicle units by different parameter sets or even by
different models.

Remarkably, the input quantities of car-following models are exactly
those of an ACC system. As in microscopic models, the ACC controller
unit calculates the acceleration with a negligible response
time. Therefore, one might state that car-following models describe
ACC systems more accurately than human drivers despite of their
intention to reproduce the traffic dynamics of human driving behavior.

Thus, the question arises, how to take into account the \textit{human}
aspects of driving for a realistic description of the traffic
dynamics. The nature of human driving is apparently more
complex. First of all, the \textit{finite reaction time} of humans
results in a delayed response towards the traffic
situation. Furthermore, human drivers have to cope with imperfect
estimation capabilities resulting in \textit{perception errors} and
\textit{limited attention spans}.  These destabilizing influences
alone would lead to a more unsafe driving and a high number of
accidents if the reaction time reached the order of the time
headway. But in day-to-day situations the contrary is observed: In
dense (not yet congested) traffic, the modal value of the time headway
distribution on German or Dutch freeways (i.e., the value where it
reaches its maximum) is around 0.9~s
\cite{Kno02-data,Tilch-TGF99,VDT}, which is of the same order of
typical reaction times \cite{green-reactionTimes}. Moreover,
single-vehicle data for German freeways \cite{Kno02-data} indicate
that some drivers even drive at headways as low as 0.3 s, which is
below the reaction time by a factor of at least 2-3 even for a very
attentive driver. For principal reasons, therefore, safe driving is
not possible in this case when considering only one vehicle in front.

This suggests that human drivers achieve additional stability and
safety by scanning the traffic situation {\it several vehicles ahead}
and by {\it anticipating} future traffic situations. The question is, how
this behavior affects the overall driving behavior and performance
with respect to ACC-like driving mimicked by car-following models. Do
the stabilizing effects (such as anticipation) or the destabilizing
effects (such as reaction times and estimation errors) dominate, or do they
effectively cancel out each other? The \textit{human driver model}
(HDM) \cite{HDM} extends the car-following modeling approach by
explicitly taking into account reaction times, perception errors,
spatial anticipation (more than one vehicle ahead) and temporal
anticipation (extrapolating the future traffic situation). It turns out
that the destabilizing effects of reaction times and estimation errors
can be compensated for by spatial and temporal anticipation
\cite{HDM}. One obtains essentially the same longitudinal dynamics,
which explains the good performance of the simpler, ACC-like
car-following models.

Thus, for the sake of simplicity, we model both automated ACC-driving
and human driving with the same microscopic traffic model, but
differentiate the driving strategies by different parameter sets.

%%%%%%%%%%%%%%%%%%%%%%%%%%%%%%%%%%%%%%%%%%%%%%%%%%%%
%%%%%%%%%%%%%%%%%%%%%%%%%%%%%%%%%%%%%%%%%%%%%%%%%%%%
%
\section{\label{sec:VLA}'Jam-avoiding' ACC driving strategies}
As discussed in the previous section, both human drivers and
ACC-controlled vehicles are effectively described by the car-following
model approach. Here, we will use the {\it intelligent
driver model} (IDM) \cite{Opus}, according to which the acceleration of each vehicle
$\alpha$ is a continuous function of the velocity $v_{\alpha}$, the
net distance gap $s_{\alpha}$, and the velocity difference
(approaching rate) $\Delta v_{\alpha}$ to the leading vehicle:
\begin{equation}
\label{IDMv}
\dot{v}_{\alpha} = a
         \left[ 1 -\left( \frac{v_{\alpha}}{v_0} \right)^4 -\left(
         \frac{s^*(v_{\alpha},\Delta v_{\alpha})} {s_{\alpha}}
         \right)^2 \right].
\end{equation}
The deceleration term depends on the ratio between the effective
'desired minimum gap'
\begin{equation}\label{sstar}
s^*(v, \Delta v) = s_0 + v T + \frac{v \Delta v } {2\sqrt{a b}}
\end{equation}
and the actual gap $s_\alpha$.  The minimum distance $s_0$ in
congested traffic is significant for low velocities only.  The
dominating term in stationary traffic is $vT$, which corresponds to
following the leading vehicle with a constant safe time headway $T$.
The last term is only active in non-stationary traffic and implements
an accident-free, 'intelligent' driving behavior including a braking
strategy that, in nearly all situations, limits braking decelerations
to the 'comfortable deceleration' $b$.  The IDM guarantees crash-free
driving. The parameters for the simulations are given in
Table~\ref{tab:IDM}.

In order to design a jam-avoiding behavior for the ACC vehicles, we
modify the ACC model parameters. The (average) time headway has a
direct relation to the maximum (static) road capacity: Neglecting the
length of vehicles leads to the approximative relationship $Q \approx
1/T$ between the flow $Q$ and the headway $T$ (cf. Eq.~(\ref{eq:qmax})
below). The crucial parameter controlling the capacity is, therefore,
the safe time headway, which is an explicit parameter of the IDM.
Moreover, the system performance is not only determined by the time
headway distribution, but also depends on the {\it stability} of
traffic flow. An ACC driving behavior aiming at increasing the traffic
performance should, therefore, additionally consider a driving
strategy which is able to stabilize the traffic flow, e.g. by a faster
dynamic adaptation to the new traffic situation.  The stability is
mainly affected by the IDM parameters 'maximum acceleration' and
'desired deceleration', see \cite{Opus}.

In the following, we will investigate the potentials of three
different parameter sets for jam-avoiding driving behavior, varying
the IDM parameters $T$, $a$ and $b$. In order to refer to the values
given in Table~\ref{tab:IDM}, we express the parameter changes by
simple multipliers. For example, $\lambda_a=2$ represents an increased
ACC parameter $a'=\lambda_a a$, where $a$ is the value listed in
Table~\ref{tab:IDM}.
\begin{enumerate}
\item[(1)]
The reduction of the time headway $T$ by a factor $\lambda_T=2/3$ has
a positive impact on the capacity. The other model parameters of
Table~\ref{tab:IDM} remain unchanged, i.e., in particular,
$\lambda_a=1,\,\lambda_b=1$.
\item[(2)]
Besides setting $\lambda_T=2/3$, we increase the desired acceleration
by choosing $\lambda_a=2$. The faster acceleration towards the desired velocity
increases the traffic stability.
\item[(3)]
The additional reduction of the desired deceleration by $\lambda_b=1/2$
corresponds to a more cautious and more anticipative driving
style. This behavior also increases the stability. 
\end{enumerate}

%##################################################################
\begin{table}
\begin{center}
\begin{tabular}{ll}
Model Parameter                                 & Value \\ \hline
Desired velocity $v_0$                          & 120 km/h \\
Save time headway $T$                           & 1.5 s \\
Maximum acceleration $a$                        & 1.0 m/s$^2$ \\
Desired deceleration $b$                        & 2.0 m/s$^2$ \\
Jam distance $s_0$                              & 2 m\\
\end{tabular}

 \caption{\label{tab:IDM}Model parameters of the \textit{intelligent
 driver model} (IDM) used in our simulations. The vehicle length is
 5~m. In order to model 'jam-avoiding' ACC strategies, we modify the
 safe time headway parameter $T$, the 'maximum acceleration' $a$ and
 the 'desired deceleration' $b$ by multipliers $\lambda_T$,
 $\lambda_a$, and $\lambda_b$, respectively.}

\end{center}
\end{table}
%##################################################################

%
%%%%%%%%%%%%%%%%%%%%%%%%%%%%%%%%%%%%%%%%%%%%%%%%%%%%
%
%%%%%%%%%%%%%%%%%%%%%%%%%%%%%%%%%%%%%%%%%%%%%%%%%%%%
%
\section{\label{sec:results}Microscopic simulations of mixed traffic}
Let us now investigate the impact of ACC vehicles which are designed
to enhance the capacity and stability of traffic flows. We will simulate
mixed traffic consisting of human and automated (ACC) longitudinal
control with a variable percentage of ACC vehicles.

Our simulation is carried out a single-lane road with an on-ramp
serving as bottleneck and with open boundary conditions. To keep
matters simple, we replace an explicit modeling of the merging of ramp
vehicles to the main road by inserting ramp vehicles centrally into
the largest gap within a $300$~m long ramp section.  In order to
generate a sufficient velocity perturbation in the merge area, the
speed of the accelerating on-ramp vehicles at the time of insertion is
assumed to be 50\% of the velocity of the respective front vehicle.

Moreover, we neglect trucks and multi-lane effects. While these
aspects are relevant in real traffic, they do not change the picture
qualitatively. Nevertheless, the induction of a second driver-vehicle
type, e.g., ACC vehicles, always has the potential to reduce the
traffic performance by an increased level of heterogeneity. We have
compared the simulation results with Gaussian distributed model
parameters, but found no qualitative difference for this single-lane
scenario.

%%%%%%%%%%%%%%%%%%%%%%%%%%%%%%%%%%%%%%%%%%%%%%%%%%%%
%
%%%%%%%%%%%%%%%%%%%%%%%%%%%%%%%%%%%%%%%%%%%%%%%%%%%%
\subsection{\label{subsec:demo}Spatiotemporal dynamics and travel time}
Let us now demonstrate that already a moderate increase in the dynamic
capacity obtained by a small percentage of 'jam-avoiding' ACC vehicles
may have a significant effect on the system performance.

We have simulated idealized rush-hour conditions by linearly
increasing the inflow at the upstream boundary over a period of 2
hours from 1200 vehicles/h to 1600 vehicles/h. Afterwards, we have
linearly decreased the traffic volume to 1000 vehicles/h until
$t=5\,$h. Moreover, we have assumed a constant ramp flow of 280
vehicles/h. Since the maximum overall flow of 1880 vehicles/h exceeds
the road capacity, a traffic breakdown is provoked at the
bottleneck. We have used the IDM parameters from Table~\ref{tab:IDM}
and parameter set (3) for ACC vehicles, i.e., $\lambda_T=2/3,
\lambda_a=2, \lambda_b=1/2$.

\begin{figure}
\begin{center}
\begin{tabular}{cc}
% bitmapped diagrams --> smaller size
  \includegraphics[width=60mm]{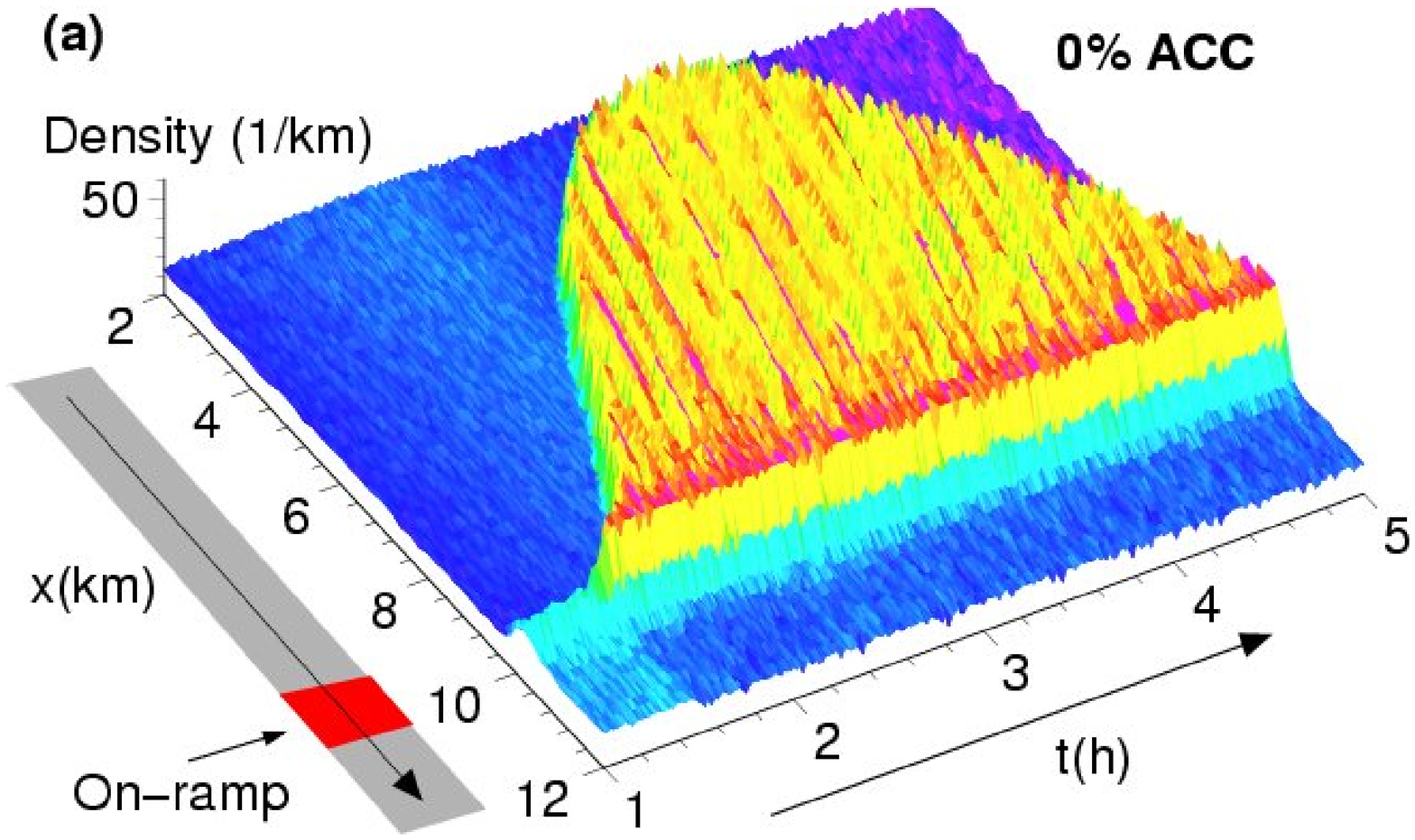}	&
  \includegraphics[width=60mm]{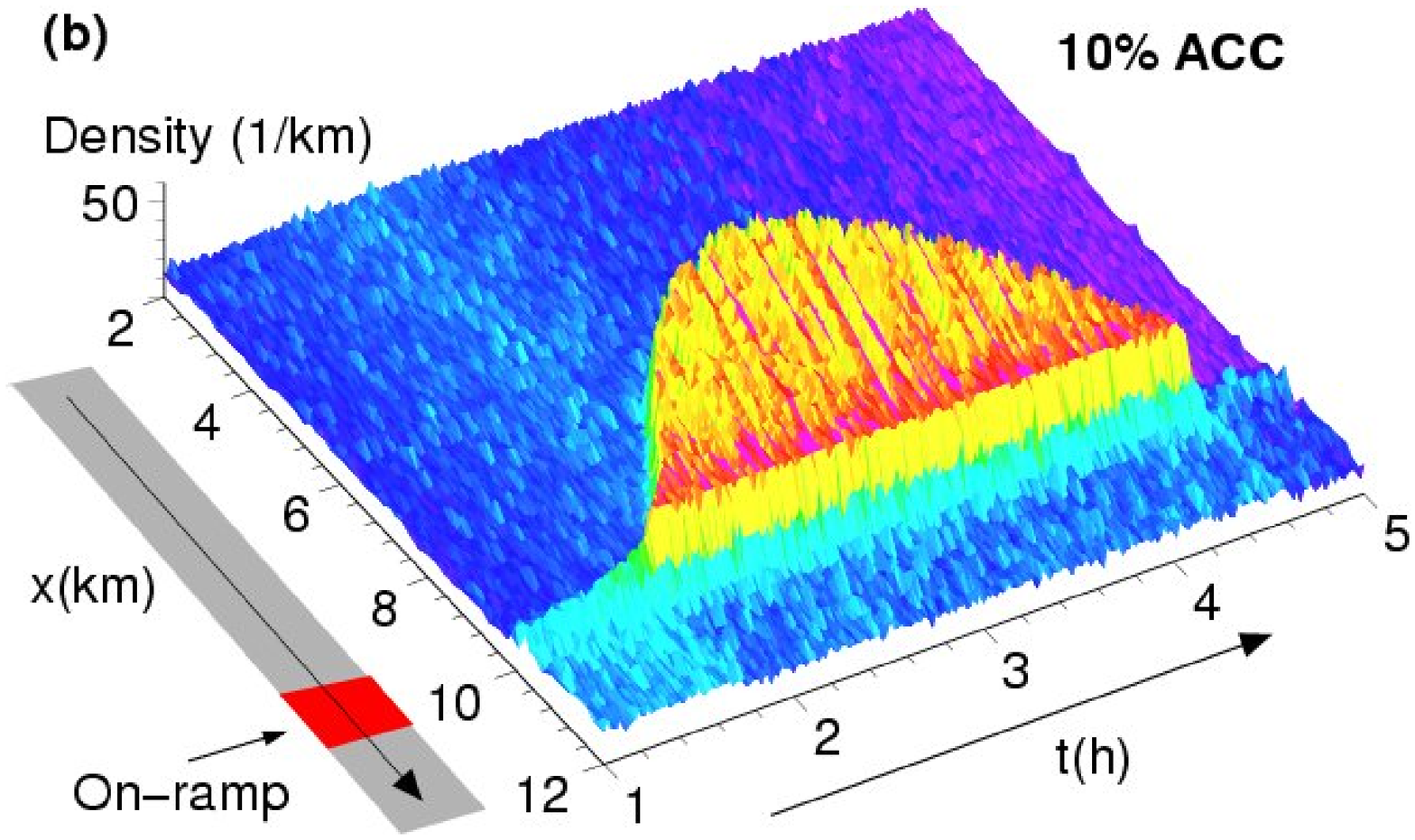}
\end{tabular}
\end{center}

 \caption{\label{fig:3d}Spatiotemporal dynamics of the traffic density
 (a) without ACC vehicles and (b) with $10\%$ ACC vehicles (parameter
 set (3)). Already a small increase in the road capacity induced by a
 small percentage of 'jam-avoiding' ACC vehicles leads to a
 significant reduction of traffic congestion (light high-density
 area).}

\end{figure}

Figure~\ref{fig:3d} shows the spatiotemporal dynamics of the traffic
density for 0\% and 10\% ACC vehicles. The increased capacity obtained
by the induced ACC vehicles leads to a strong reduction of the traffic
jam already for a small percentage of ACC vehicles. For 30\% ACC
vehicles, the traffic jam disappears completely.

\begin{figure}
\begin{center}
\begin{tabular}{cc}
	\includegraphics[width=60mm]{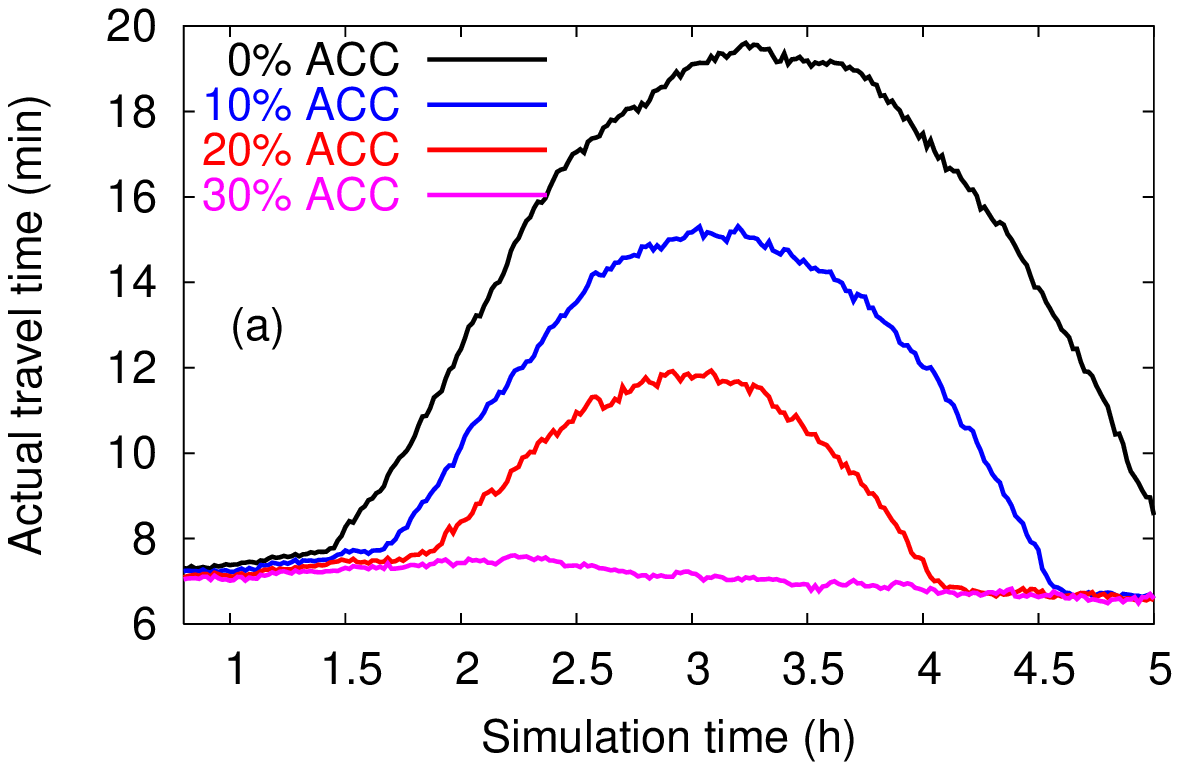}	&
	\includegraphics[width=60mm]{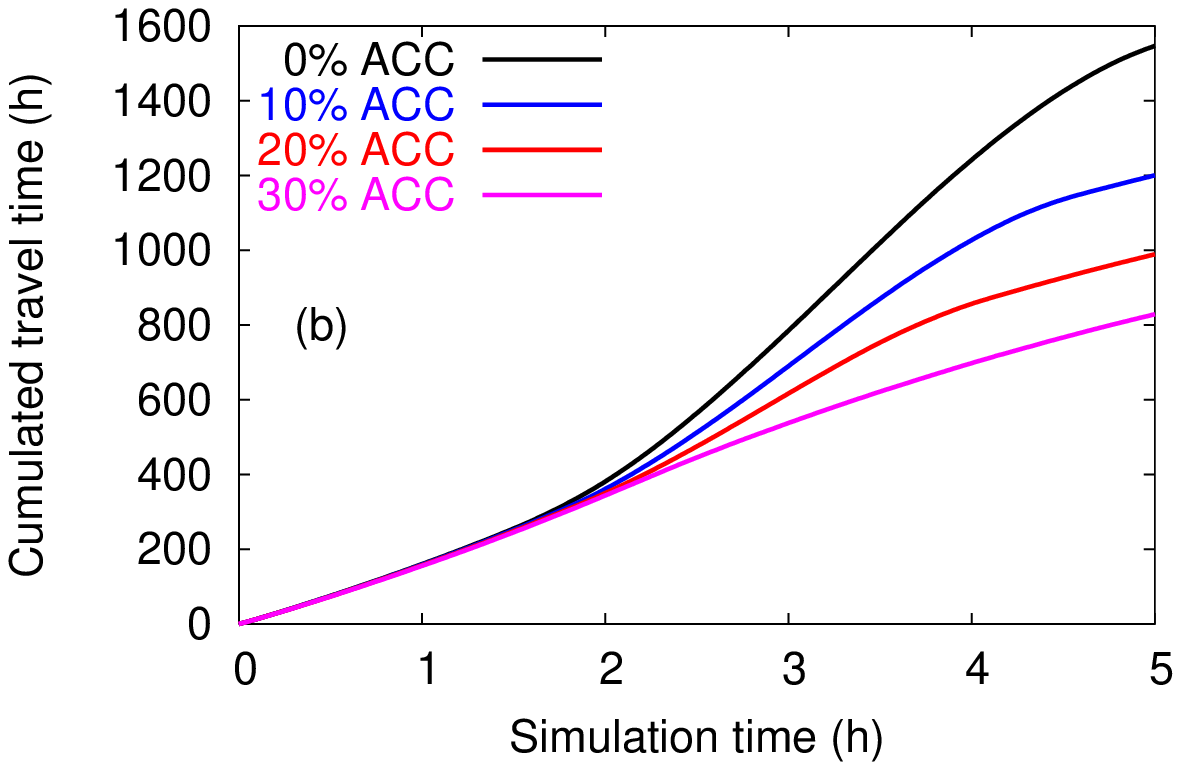}
\end{tabular}
\end{center}

 \caption{\label{fig:tt}Time series for (a) the actual and (b) the
 cumulated travel times for simulation runs with different percentages
 of ACC vehicles. The traffic breakdown leads to a significant
 prolongation of travel time.  A proportion of 30\% ACC vehicles can
 completely prevent the traffic breakdown.}

\end{figure}

An increased percentage of 'jam-avoiding' ACC vehicles has a strong
effect on the travel time: Figure~\ref{fig:tt} shows the actual and
cumulated travel times for various ACC percentages. At the peak of
congestion ($t=3.2\,$h), the travel time for individual drivers is
nearly triple that of the uncongested situation ($t<1\,$h). Already
10\% ACC vehicles reduce the maximum travel time delay of individual
drivers by about 30\% (Fig.~\ref{fig:tt}(a)), and the cumulated time
delay (which can be associated with the economic cost of this jam) by
50\% (Fig.~\ref{fig:tt}(b)). Several factors contribute to this
enhanced system performance. First, an increased ACC percentage leads
to a delay of the traffic breakdown.  Second, the ACC vehicles reduce
the maximum queue length significantly. Third, the jam dissolves
earlier. These effects, which are responsible for the drastic increase
in the system performance already for a small proportion of 'jam-avoiding'
ACC vehicles, will be investigated in the following.

%##################################################################
%##################################################################

%##################################################################
%##################################################################

%%%%%%%%%%%%%%%%%%%%%%%%%%%%%%%%%%%%%%%%%%%%%%%%%%%%
% free flow
%%%%%%%%%%%%%%%%%%%%%%%%%%%%%%%%%%%%%%%%%%%%%%%%%%%%
\subsection{\label{subsec:free}Maximum capacity in free traffic}
The \textit{static road capacity} $Q_\text{max}^\text{theo}$, which
corresponds to the maximum of the flow-density diagram, is mainly
determined by the average time headway $T$. However, the theoretical capacity
depends also on the 'effective' length $l_\text{eff}=l_\text{veh}+s_0$ of
a driver-vehicle unit and is given by
\begin{equation}\label{eq:qmax}
Q_\text{max}^\text{theo} = \frac{1}{T} \left(1 - \frac{l_\text{eff}}{v_0 T+l_\text{eff}} \right).
\end{equation}
The maximum capacity $Q_\text{max}^\text{free}$ before traffic breaks
down (which is a \textit{dynamic} quantity), however, is typically
lower than $Q_\text{max}^\text{theo}$, since it depends on the traffic
stability as well. Therefore, we have analyzed the 'maximum free
capacity' resulting from the traffic dynamics as a function of the
average time headway $T$ and the percentage of ACC vehicles. Our
related simulation runs start with a low upstream inflow and linearly
increase the inflow with a rate of
$\dot{Q}_\text{in}=800$~vehicles/h$^2$. We have checked other
progression rates as well, but found a marginal difference only.

\begin{figure}
\begin{center}
  \begin{tabular}{cc}
  \includegraphics[width=0.5\textwidth]{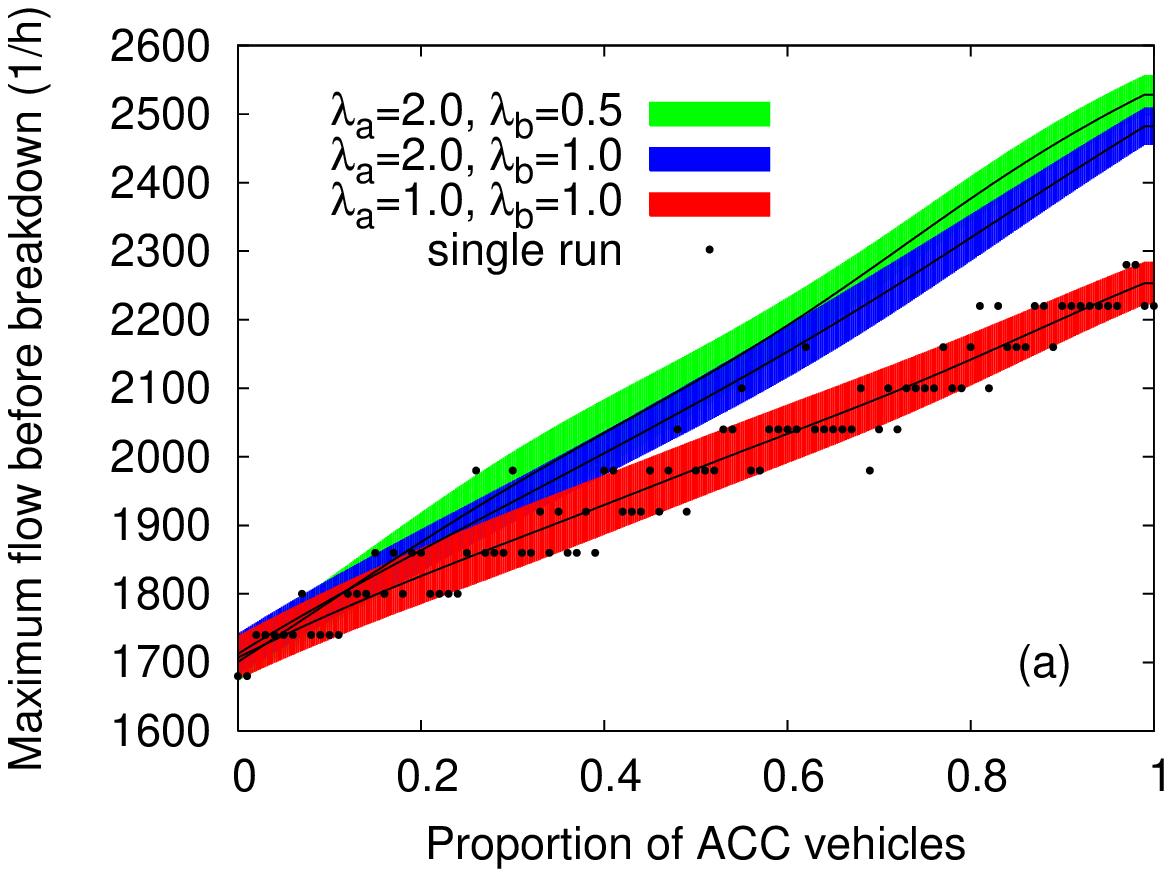}  &
  \includegraphics[width=0.5\textwidth]{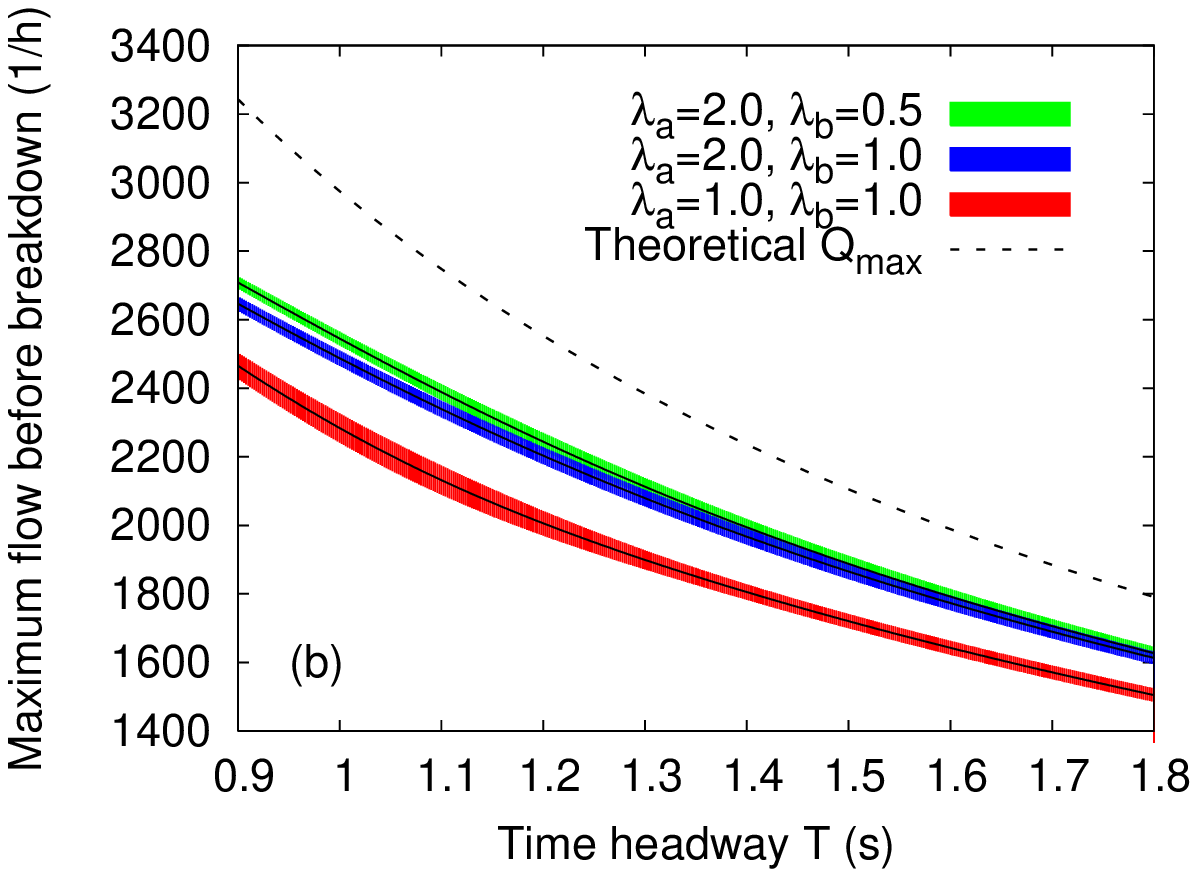}
\end{tabular}
\end{center}

 \caption{\label{fig:free}Maximum free capacity as a function of (a)
 the percentage of ACC vehicles, and (b) the time headway $T$ for
 100\% ACC vehicles. We have simulated three different parameter sets
 for ACC vehicles with $\lambda_T=2/3$ and varying values of
 $\lambda_a$ and $\lambda_b$ (see main text).  Dots indicate results
 of single simulation runs, while the solid lines correspond to
 averages over several simulations and the associated bands to
 plus/minus one standard deviation.}

\end{figure}

For determining the traffic breakdown, we have used 'virtual
detectors' located $1\,$km upstream and downstream of the on-ramp
location. In analogy to the real-world double-loop detectors, 'virtual
detectors' count the passing vehicles, measure the velocities, and
aggregate the data within a time interval of one minute. For each
simulation run, we have recorded the maximum flow before traffic has
broken down (single dots in Fig.~\ref{fig:free}(a)). Due to the
complexity of the simulation and the 1-min data aggregation,
$Q_\text{max}^\text{free}$ varies stochastically. We have, therefore,
averaged the data with a linear regression using a Gaussian weight of
width $\sigma=0.2$, and plotted the expectation value and the standard
deviation.

Figure~\ref{fig:free}(a) shows the maximum free capacity as a function
of the ACC percentage for the three different parameter sets
representing different ACC driving styles. $Q_\text{max}^\text{free}$
increases approximately linearly with increasing percentage of ACC
vehicles. The parameter $a$ mainly increases the traffic stability,
which leads to a delayed traffic breakdown and, thus, to higher values
of $Q_\text{max}^\text{free}$. Remarkably, the values are nearly
identical with those for heterogenous traffic consisting of
driver-vehicle units with Gaussian distributed parameters.

In Fig.~\ref{fig:free}(b) the most important parameter, the time
headway $T$, is varied for a homogeneous ensemble of 100\% ACC
vehicles. Obviously, $Q_\text{max}^\text{free}$ decreases with
increasing $T$. Furthermore, the dynamic quantity
$Q_\text{max}^\text{free}$ remains always lower than the theoretical
capacity $Q_\text{max}^\text{theo}$ given by Eq.~(\ref{eq:qmax}), which
is only reached for perfectly stable traffic. The three parameter sets
show the influence of the IDM parameters $a$ and $b$: The acceleration
$a$ has a strong impact on traffic stability, while the stabilizing
influence of $b$ is smaller. Finally, as the difference between
$Q_\text{max}^\text{theo}$ and the dynamic maximum free capacity
$Q_\text{max}^\text{free}$ increases for lower values of $T$, one finds
that a smaller $T$ reduces stability as well.

In order the assess the potentials of various driving styles, we have
evaluated an approximate relationship as a function of the ACC
equipment level $\alpha_\text{ACC}$. The relative gain $\gamma$ in
system performance is given by
\begin{equation}
\gamma \approx \left[0.95(1-\lambda_T)+0.07 \lambda_a+0.08 (1-\lambda_b)\right] \,\alpha_\text{ACC}.
\end{equation}
Thus, $\lambda_T$ is the most crucial parameter, while $\lambda_b$ has
hardly any influence. 
For example, lowering the time headway by $\lambda_T=0.7$ with
$\alpha_\text{ACC}=1$ results in a maximum gain of $\gamma\approx 30\%$.

%##################################################################
%##################################################################

%%%%%%%%%%%%%%%%%%%%%%%%%%%%%%%%%%%%%%%%%%%%%%%%%%%%
%
%%%%%%%%%%%%%%%%%%%%%%%%%%%%%%%%%%%%%%%%%%%%%%%%%%%%
%
\subsection{\label{subsec:dynC}Dynamic capacity after a traffic breakdown}
Let us now investigate the system dynamics after a traffic
breakdown. The crucial quantity is the \textit{dynamic capacity},
i.e., the downstream outflow from a traffic congestion $Q_\text{out}$
\cite{Daganzo}. The difference between the
free capacity $Q_\text{max}^\text{free}$ and $Q_\text{out}$ is denoted
as \textit{capacity drop} with typical values between 5\% and 30\%.

We have used the same simulation setup as in the previous
section. After a traffic breakdown was provoked by an increasing
inflow, we have averaged over the 1-min flow data of the 'virtual
detector' $1\,$km downstream of the bottleneck. We have identified the
congested traffic state by filtering out for velocities smaller than
$50\,$km/h at a cross-section $1\,$km upstream of the
bottleneck. Again, we have averaged over multiple simulation runs by
applying a Gaussian-weighted linear regression.

\begin{figure}
\begin{center}
  \begin{tabular}{cc}
	\includegraphics[width=60mm]{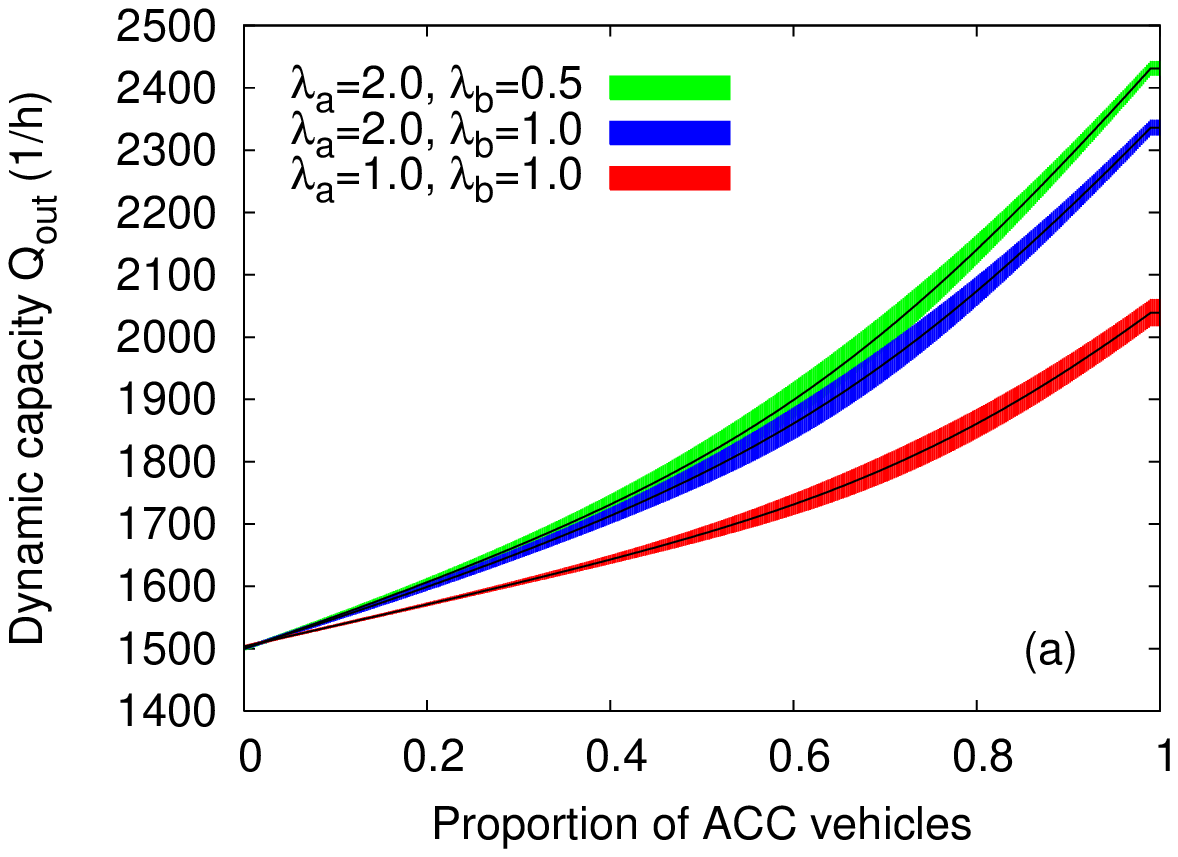} &
	\includegraphics[width=60mm]{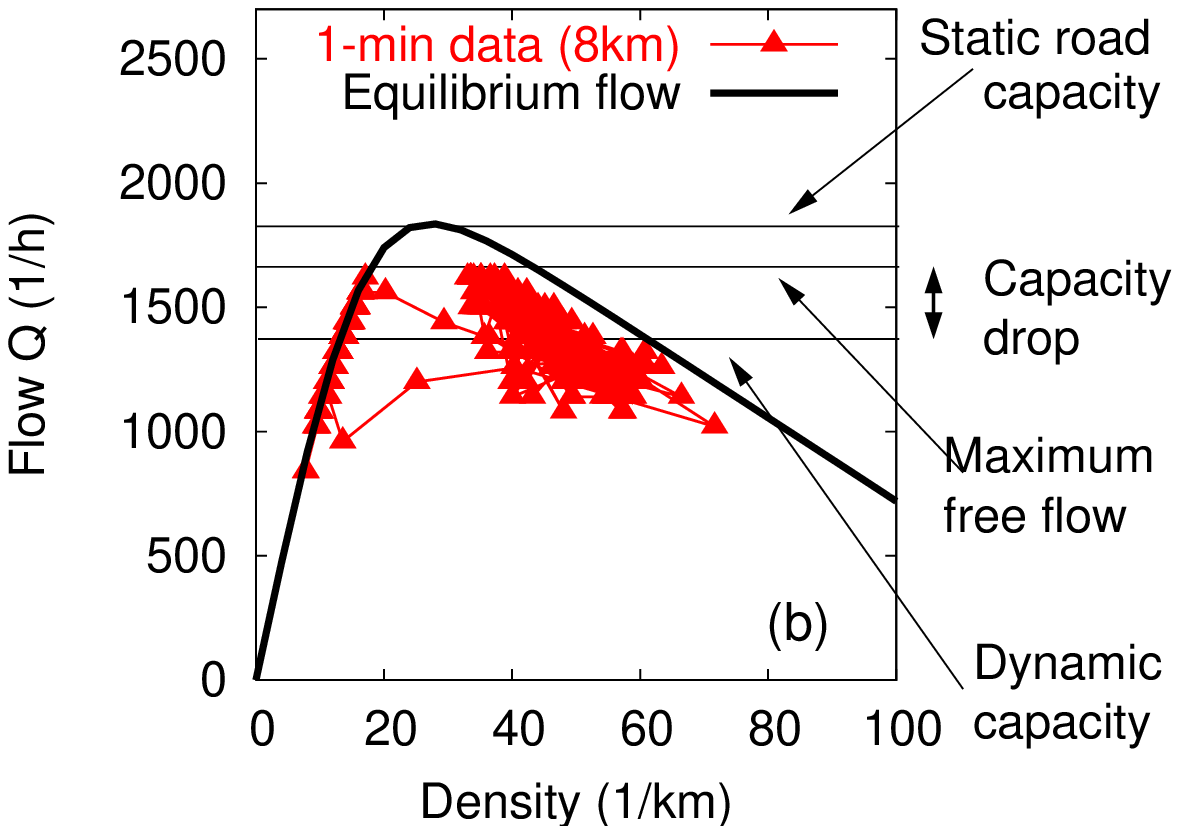}
\end{tabular}
\end{center}

 \caption{\label{fig:q_out}(a) Dynamic capacity as a function of the
 percentage of ACC vehicles. The curves represent three different
 parameter sets corresponding to different ACC driving strategies. The
 results from multiple simulation runs are averaged using a linear
 regression with a Gaussian weight of width $\sigma=0.2$. (b)
 Flow-density data for the traffic breakdown determined from a
 'virtual' detector 2 km upstream of the bottleneck without ACC
 vehicles. The equilibrium flow-density curve of identical vehicles
 corresponds to the parameter set given in
 Table~\protect\ref{tab:IDM}.}

\end{figure}

Figure~\ref{fig:q_out}(a) shows the dynamic capacity for a variable
percentage of ACC vehicles for the three different parameter sets
specified before.  Interestingly, the capacity increase is not linear
as in Fig.~\ref{fig:free}(a). Above approximately 50\% ACC vehicles,
the dynamic capacity increases faster than for lower percentages. We
explain this behavior with an 'obstruction effect': the faster
accelerating ACC vehicles are hindered by the slower accelerating
drivers. In fact, the slowest vehicle type determines the dynamic
capacity, which could be called a 'weakest link effect'. In
conclusion, distributed model parameters have a quantitative effect on
the outflow from congested traffic (it is lower than for homogeneous
traffic with averaged parameters), while such an effect is not
observed for the free-flow capacity!

%##################################################################
%##################################################################

%%%%%%%%%%%%%%%%%%%%%%%%%%%%%%%%%%%%%%%%%%%%%%%%%%%%
%
%%%%%%%%%%%%%%%%%%%%%%%%%%%%%%%%%%%%%%%%%%%%%%%%%%%%
%
\section{\label{sec:diss}Discussion}
Adaptive cruise control (ACC) systems are already available on the
market. The next generations of ACC systems will extend their range of
applicability to all speeds, and it is assumed that their spreading
will grow in the future. In this contribution, by means of microscopic
traffic simulations we have investigated the impact that an automated
longitudinal driving control of ACC systems based on the intelligent
driver model (IDM) is expected to have on the traffic dynamics.

ACC systems are closely related to car-following models as their
reaction is restricted to a leading vehicle.  Moreover, we have
explained why such a car-following approach also captures the main
aspects of longitudinal driver behavior so well. We, therefore, expect
that both ACC systems and human driver behavior will mix consistently
in future traffic flows although the driving operation is
fundamentally different.

The equipment level of ACC systems provides an interesting option to
enhance the traffic performance by automated driving strategies. In
order to analyze the potentials, we have studied ACC driving styles,
which are explicitly designed to increase the capacity and stability
of traffic flows. We have varied the percentage of ACC vehicles and
found that already a small proportion of ACC vehicles, which implies a
marginally increased free and dynamic capacity, leads to a drastic
reduction of traffic congestion. Furthermore, we have shown that,
capacity and stability do have similar importance for the traffic
dynamics.

We have assumed that the ACC systems have a more 'jam-avoiding'
driving style than the human drivers. One might additionally take into
account inefficient human behavior when traffic gets denser and the
time headway increases with increasing local velocity variance
\cite{VDT,VDT-TGF05}. In this case, a constant time headway policy for
automated driving is expected to improve the system performance even
more.

Up to now, ACC systems are only optimized for the user's driving
comfort and safety. In fact, present ACC systems may have a negative
influence on the system performance when their percentage becomes
large. The design of ACC strategies, which also consider their impact
on traffic dynamics, will be crucial for the next ACC generations.

Furthermore, we propose to implement an 'intelligent' ACC strategy
that adapts the ACC driving style
\textit{dynamically} to the overall traffic situation. For example, in
dense, but not yet congested traffic, a jam-avoiding parameter set
could help to delay or suppress traffic breakdowns as shown in our
simulations, while in free traffic a parameter set mimicking natural
driver behavior may be applied instead.  The respective 'traffic
state' could be autonomously detected by the vehicles using the
history of their sensor data in combination with digital
maps. Moreover, inter-vehicle communication could contribute
information about the traffic situation in the neigborhood, e.g., by
detecting the downstream front of a traffic jam \cite{IVC-TGF05}.

\textbf{Acknowledgments:}
The authors would like to thank Hans-J\"urgen Stauss, and Klaus Rieck
for the excellent collaboration and the Volkswagen AG for partial
financial support within the BMBF project INVENT.

%##########################################################
%References
%##########################################################
\bibliographystyle{osa}
%\bibliography{/home/library/bibtex/database}

\end{document}